\documentclass[12pt]{iopart}

\usepackage{iopams}
\expandafter\let\csname equation*\endcsname\relax
\expandafter\let\csname endequation*\endcsname\relax
\usepackage{amsmath}
\usepackage[english]{babel}
\usepackage{graphicx}
\usepackage{cite}
\usepackage{color}
\usepackage{bbm}
\usepackage[normalem]{ulem}
\usepackage{times}
\usepackage[colorlinks=true,citecolor=cyan,linkcolor=cyan,urlcolor=cyan]{hyperref}
\usepackage[normalem]{ulem}
\usepackage{dsfont}
\usepackage{txfonts}

\newcommand{\ket}[1]{\vert #1 \rangle}
\newcommand{\bra}[1]{\langle #1 \vert}
\newcommand{\ketbra}[2]{\vert #1 \rangle \langle #2 \vert}

 \newcommand{\expval}[1]{\left< #1 \right>}

\begin{document}

\title[]{Floquet stroboscopic divisibility in non-Markovian dynamics}

\author{Victor M. Bastidas}
\address{Centre for Quantum Technologies, National University of Singapore,
3 Science Drive 2, Singapore 117543, Singapore}

\author{Thi Ha Kyaw}
\address{Centre for Quantum Technologies, National University of Singapore,
3 Science Drive 2, Singapore 117543, Singapore}

\author{Jirawat Tangpanitanon}
\address{Centre for Quantum Technologies, National University of Singapore,
3 Science Drive 2, Singapore 117543, Singapore}

\author{Guillermo Romero}
\address{Departamento de F\'isica, Universidad de Santiago de Chile (USACH), 
Avenida Ecuador 3493, 9170124, Santiago, Chile}

\author{Leong-Chuan Kwek}
\address{Centre for Quantum Technologies, National University of Singapore,
3 Science Drive 2, Singapore 117543, Singapore}
\address{MajuLab, CNRS-UNS-NUS-NTU International Joint Research Unit, UMI 3654, Singapore}
\address{Institute of Advanced Studies, Nanyang Technological University,
60 Nanyang View, Singapore 639673, Singapore}
\address{National Institute of Education, Nanyang Technological University,
1 Nanyang Walk, Singapore 637616, Singapore}

\author{Dimitris G. Angelakis}
\address{Centre for Quantum Technologies, National University of Singapore,
3 Science Drive 2, Singapore 117543, Singapore}
\address{School of Electrical and Computer Engineering, Technical University of Crete, Chania, Crete, 73100 Greece}

\begin{abstract}
We provide a general description of a time-local master equation for a system coupled to a non-Markovian reservoir based on Floquet theory. This allows us to have a divisible dynamical map at discrete times, which we refer to as \textit{Floquet stroboscopic divisibility}. We illustrate the theory by considering a harmonic oscillator coupled to both non-Markovian and Markovian baths. Our findings provide us with a theory for the exact calculation of spectral properties of time-local non-Markovian Liouvillian operators, and might shed light on the nature and existence of the steady state in non-Markovian dynamics.
\end{abstract}

\maketitle

%%%%%%%%%%%%%%%%%%%%%%%%%%%%%%%%%%%%%%%%%%%%%%%%%%%%%%%%%%%%%%%%%%%%%%%%%%%%%%%%%%%%%%%%
\section{Introduction}
\label{sec:I}
%%%%%%%%%%%%%%%%%%%%%%%%%%%%%%%%%%%%%%%%%%%%%%%%%%%%%%%%%%%%%%%%%%%%%%%%%%%%%%%%%%%%%%%%

    A general description of open quantum system dynamics has proven to be a challenging problem in quantum statistical mechanics~\cite{UWeiss}. Most of our knowledge is based on the description of system-bath interaction, where the memory effect of the bath plays a key role. When a quantum system is in contact with a memory-less (Markovian) bath, the information flows unidirectionally from the system to the bath. However, if the bath has memory (non-Markovian), the situation can change dramatically. As a matter of fact, there are several definitions of non-Markovianity in the literature~\cite{2014Plenio, 2016Breuer, 2017Vega}. For example, backflow of information can be used to quantify non-Markovianity, which motivates the BLP measure proposed by Breuer, Laine, and Piilo~\cite{2014Plenio, 2016Breuer, 2017Vega}. Another definition is based on properties of the dynamical map, $\hat{\Phi}(t,0)$, which is the propagator of the density matrix of the system, $\hat{\rho}_\mathcal{S}(t)$. Following Refs. ~\cite{2014Plenio, 2016Breuer, 2017Vega}, a dynamical map is Markovian when it is a trace preserving divisible map such that $\hat{\Phi}(t_2,0)= \hat{\Phi}(t_2,t_1) \hat{\Phi}(t_1,0)$ where $\hat{\Phi}(t_2,t_1)$ is completely positive for any $t_1, t_2 > 0$. A dynamical map satisfying this property is referred to as CP-divisible. In terms of this definition, to have non-Markovian dynamics, there must be some time $t_1$ such that $\hat{\Phi}(t_2,t_1)$ is not completely positive, which motivates the RHP measure of non-Markovianity proposed by Rivas, Huelga and Plenio~\cite{2014Plenio}. Furthermore, the absolute value of the determinant of the dynamical map can be interpreted as the volume of the accessible states at a given time, which motivates the geometrical characterization of non-Markovianity and the measure $\mathcal{N}_{V}$~\cite{2013Lorenzo}.  Interestingly, a map can be non-Markovian in the sense of the RHP, while it is Markovian according to the $\mathcal{N}_{V}$ measure. This occurs because a map can be P-divisible but not necessarily CP-divisible~\cite{2013Lorenzo}.

Recently, the dynamics of systems coupled to non-Markovian reservoirs have been the focus of active theoretical research~\cite{2014Plenio, 2016Breuer, 2017Vega,2013Lorenzo, 2008Wolf,2010Rivas, tan2010non,2016Strasberg}. This is partially driven by recent developments on quantum technologies which allow one to manipulate quantum systems with unprecedented precision and control. For instance, structured reservoirs appear naturally in the study of a driven qubit coupled to a damped detector~\cite{2005Goorden}. Also, one can use superconducting qubits to simulate structured reservoirs that are relevant for the study of exciton transport in photosynthetic complexes~\cite{2014Mostame} and Zeno effect of a single superconducting qubit coupled to an array of transmission line resonators~\cite{tong2014simulating}. By using the reaction coordinate mapping, it is possible to explore nonequilibrium thermodynamics in the non-Markovian regime~\cite{2016Strasberg}. 
A recent paper discusses an operational method to characterize arbitrary quantum processes in terms of a mapping to a manybody state with a matrix-product-state representation, which can be applied to non-Markovian dynamics~\cite{Pollock}. In addition, non-Markovian behavior has been explored in the context of photonic systems with structured reservoirs~\cite{2000Lambropoulos} that even allow one to inhibit spontaneous emission of an atom embedded in a photonic crystal~\cite{2004Angelakis}. In some situations where the reservoir is structured or under the effect of an external drive, the open-system approach is inadequate to describe the dynamics of the system and it is suitable to study the combined dynamics of the system and the environment as in Ref~\cite{2016Restrepo}. Besides the theoretical investigations, there are experimental realizations of non-Markovian dynamics in all-optical setups~\cite{2011Liu}, trapped ions~\cite{2013Schindler,2017Schaetz}, and optomechanical systems~\cite{2015Groblacher}, to mention but a few.

In the case of a system coupled to a bath, one can carry out a microscopic derivation of the master equation for the reduced density matrix of the system using the open-systems approach ~\cite{UWeiss,2002Breuer,2004gardiner}. After performing the Born, Markov, and secular approximations, the resulting master equation has a Lindblad form with positive rates and the corresponding dynamical map is CP-divisible~\cite{2014Plenio, 2016Breuer, 2017Vega}. However, as it is discussed in Ref.~\cite{2014Plenio}, this is not the only framework to obtain Markovian dynamics. Furthermore, in the case of a Lindblad-type master equation, the rates can be time-dependent, but as long as they are positive at all times, the corresponding master equation leads to a CP-divisible map. In this work we restrict ourselves to master equations in the Lindblad form~\cite{1976lindblad,1976gorini,2004gardiner}, which can be written formally in terms of a Liouvillian operator (LO) that is local in time~\cite{2014Plenio, 2016Breuer, 2017Vega}. In the case of time-independent rates, the eigenvalues of the LO are known as the Liouvillian spectrum. The imaginary and real parts of the Liouvillian spectrum are related to coherent and incoherent processes, respectively. In addition, the kernel of the LO determines the steady state of the system. If the LO is time-independent or it has an adiabatic dependence on time, one can diagonalize it to obtain its spectrum. However, this is not the case for non-adiabatic time dependence. Time dependent LOs appear when the system is driven externally or due to time-dependent damping rates. For a long time, the theoretical understanding of time dependent LOs has been an open problem~\cite{1978davies,1986Lendi,2016Dai,2016Restrepo}. These kind of LOs lead to time-local (time-convolutionless) master equations, \textit{which can be non-Markovian} when the damping rates become negative at certain times~\cite{2013Haikka,2012Fleming}. This type of master equations appear naturally in the context of pure-dephasing channels~\cite{1996Palma,2002Reina}. A recent work~\cite{2017Hartmann} has shown that for a Markovian master equation with time-periodic LO, one can use Floquet theory~\cite{1883floquet, 1975yakubovich,1998GRIFONI,1997Kohler} to obtain the asymptotic steady state.

In this article we use the Floquet theory to generalize the definition of the Liouvillian spectrum to non-Markovian dynamics. The latter is generated through a time-periodic LO in Lindblad form, such that the system dynamics is ruled by the equation $\frac{d \hat{\rho}_\mathcal{S}(t)}{dt}=\hat{\mathcal{L}}(t)\hat{\rho}_\mathcal{S}(t)$, where $\hat{\mathcal{L}}(t+T)=\hat{\mathcal{L}}(t)$. The non-Markovianity is guaranteed by periodic damping rates which are negative in certain time intervals. Although in general the dynamics is not P-divisible, the Floquet theorem ensures that there exists a dynamical map $\hat{\Phi}(t;0)=\hat{P}(t)\exp(\hat{\mathcal{L}}_{\text F}t)$, where $\hat{P}(t+T)=\hat{P}(t)$~\cite{1883floquet, 1975yakubovich}. In this case, it is direct to prove that the dynamical map is divisible at discrete times, $\hat{\Phi}(mT;0)=[\hat{\Phi}(T;0)]^m$ with integer $m$. In addition, if the map is completely positive, it is also CP-divisible at stroboscopic times, which we term as \textit{Floquet stroboscopic divisibility}. 
Importantly, the eigenvalues of the matrix $\hat{\Phi}(T;0)$ allow us to fully characterize the spectral properties of the non-Markovian LO. We illustrate this theory by considering a quantum harmonic oscillator coupled to two dephasing baths: one is non-Markovian and another is Markovian. This leads to constant and time-periodic dephasing rates, from the Markovian and non-Markovian baths, respectively. We observe that the dynamics undergoes a transition from Markovian to non-Markovian behavior as the coupling to the non-Markovian bath is increased.  
Our findings might shed light on the nature and existence of the steady state in non-Markovian dynamics.

%%%%%%%%%%%%%%%%%%%%%%%%%%%%%%%%%%%%%%%%%%%%%%%%%%%%%%%%%%%%%%%%%%%%%%%%%%%%%%%%%%%%%%%%
\section{Floquet stroboscopic divisibility}
\label{sec:II}
To make a direct connection between the dynamics of an open quantum system  and Floquet theory, we consider a time-local~\cite{1976lindblad,1976gorini,2017Hartmann} master equation $\frac{d \hat{\rho}_{S}(t)}{d t}=\hat{\mathcal{L}}(t)\hat{\rho}_\mathcal{S}(t)$ with time periodic LO $\hat{\mathcal{L}}(t+T)=\hat{\mathcal{L}}(t)$. Here, $\hat{\rho}_\mathcal{S} (t)$ denotes the reduced density matrix of the system. One can define a propagator $\hat{\Phi}(t;0)$, or dynamical map, such that $\hat{\rho}_\mathcal{S}(t)=\hat{\Phi}(t;0)\hat{\rho}_\mathcal{S}(0)$. 
Due to the periodic nature of the LO, the dynamical map satisfies the condition $\hat{\Phi}(lT;0)=[\hat{\Phi}(T;0)]^l$ with integer $l$. If we take $l=m+n$ in the previous identity, one can show that the map is divisible at stroboscopic times, i.e., $\hat{\Phi}[(m+n)T;0]=\hat{\Phi}(mT;0)\hat{\Phi}(nT;0)$. If  $\hat{\Phi}(T;0)$ is not only positive, but completely positive~\cite{2014Plenio, 2016Breuer, 2017Vega}, then from the previous identities it follows that the map is also CP-divisible at stroboscopic times, which we term as \textit{Floquet stroboscopic divisibility}.

Our aim now is to interpret the dynamics in terms of Floquet theory~\cite{1883floquet, 1975yakubovich}. Given a basis for the system Hilbert space, the master equation turns out to be just a system of coupled ordinary differential equations with periodic coefficients. For example, if one represents the density matrix $\hat{\rho}_{\text{S}}(t)$ as a vector,  the matrix representation of the Liouvillian will be  a time-periodic matrix. This allows us to apply the Floquet theorem for ordinary differential equations with periodic coefficients~\cite{1883floquet, 1975yakubovich}.  The Floquet theorem ensures that there exists a dynamical map $\hat{\Phi}(t;0)$\textemdash or fundamental matrix\textemdash with the form $\hat{\Phi}(t;0)=\hat{P}(t)\exp(\hat{\mathcal{L}}_{\text F}t)$, where $\hat{P}(t+T)=\hat{P}(t)$~\cite{1883floquet, 1975yakubovich}. The eigenvalues $\lambda_{\alpha}=e^{L_{\alpha}T}$ of the matrix $\hat{\Phi}(T;0)$ and the complex eigenvalues $L_{\alpha}\in \mathbb{C}$ of $\hat{\mathcal{L}}_{\text{F}}$ are called the characteristic multipliers and the \textit{Floquet exponents} (Floquet-Liouville spectrum), respectively. Furthermore, the Floquet theorem provides us with a suggestive form $\hat{\Phi}(lT;0)=\exp(\hat{\mathcal{L}}_{\text F}\ lT)$ of the dynamical map at stroboscopic times $t=lT$, which resembles the dynamical map in the case of a time-independent LO.

In a similar way that for time-independent LOs, the imaginary part of the spectrum governs the coherent dynamics, and the real part is responsible for incoherent/dissipative processes. So far we have discussed spectral properties of the dynamical map, but the Floquet theorem gives us more information. For example, the solution of the master equation can be written as $\rho_{\text{S}}(t)=\sum_{\alpha}c_{\alpha}e^{L_{\alpha}t}\rho_{\alpha}(t)$, where $\rho_{\alpha}(t+T)=\rho_{\alpha}(t)$ and $\hat{\Phi}(T)\rho_{\alpha}(T)=e^{L_{\alpha}T}\rho_{\alpha}(T)$. One should also take into account that the Floquet exponents are not uniquely defined because one can always add a complex phase $2\pi n i/T$ with integer $n$ such that one gets the same characteristic multiplier, i.e, $e^{(L_{\alpha}+2\pi n i/T)T}=e^{L_{\alpha}T}$~\cite{1883floquet, 1975yakubovich}. The kernel of the operator  $\hat{\mathcal{L}}_{\text{F}}$ is a solution of the equation  $\hat{\Phi}(T)\rho_{\alpha}(T)=\rho_{\alpha}(T)$ and determines the steady state. In the case of a Lindblad-type master equation with time-dependent decay rates that are always positive, the dynamical map is CP-divisible~\cite{2014Plenio, 2016Breuer, 2017Vega} and the dynamics are Markovian. For positive time-periodic decay rates, the Floquet theorem ensures the existence of a periodic steady state as it is shown in Ref.~\cite{2017Hartmann}. In contrast, in the non-Markovian case, the existence of a steady state is highly nontrivial as it is discussed in Ref.\cite{2017Vega}.

\textit{Floquet stroboscopic divisibility} is a direct consequence of the Floquet theorem because at stroboscopic times, the dynamical map is CP-divisible. But this alone is not enough to ensure the existence of a steady state, because we still need to prove that the dynamical map is a contractive map at stroboscopic times. With this aim, we need to resort in spectral properties of the dynamical map. 
The Floquet theorem establishes that stable solutions are possible when the Lyapunov exponents, i.e., the real part of the Floquet exponents, are smaller than or equal to zero~\cite{1883floquet, 1975yakubovich}. That implies the stability constraint $|\det\hat{\Phi}(T;0)|\leq 1$, which can be derived from the general formula $\det\hat{\Phi}(t;0)=\exp\{\int_0^{t}\text{Tr}[\hat{\mathcal{L}}(\tau)]d\tau\}$ see Refs.~\cite{1883floquet, 1975yakubovich}.
The absolute value of the determinant of the dynamical map can be reinterpreted as the volume of the accessible states at a given time, which motivates the geometrical characterization of non-Markovianity~\cite{2013Lorenzo}. Within this framework, if a dynamical map is P-divisible then the rate of change of the volume of available states is smaller than zero~\cite{2014Maniscalco,2017Chruscinski}. 

We are interested in the case where the time average of all the rates in one period is positive or zero, in order to satisfy the stability constraint. We also note that the previous statement does not restrict the rates to be positive at all times. 
In contrast to Refs.~\cite{2014Maniscalco,2017Chruscinski}, we need to define the rate of change of the volume of available states in a discrete way, due to the stroboscopic nature of the evolution. In our case, the dynamical map is stroboscopically contractive if the finite differences $\Delta_m=\frac{1}{T}(|\det\hat{\Phi}[(m+1)T]|-|\det\hat{\Phi}(mT)|)$ satisfy the condition
%%%
\begin{equation}
      \label{eq:stroboscopicVol}
            \Delta_m=\frac{|\det\hat{\Phi}(T)|^m}{T}(|\det\hat{\Phi}(T)|-1)\leq 0
      \ .
\end{equation}
%%%
Interestingly enough, $\Delta_m$ goes to zero either when $m$ goes to infinity and the determinant is smaller than one, or when the determinant is equal to one for any $m$. In the former case, this implies that the system reaches a periodic asymptotic state. The latter means that the system is purified stroboscopically. In contrast to the results presented in Refs.~\cite{2013Lorenzo,2014Maniscalco,2017Chruscinski}, $\Delta_m$ is a measure of how the volume of accessible states is contracted stroboscopically.  
In the following, we will apply the general theory presented so far to a simple example: a harmonic oscillator couples to both Markovian and non-Markovian baths, see figure \ref{fig:fig0}(a). This is one example of the general theory, but there are other possible examples such as phase- and amplitude-damped qubits~\cite{2014Addis}. Based on our example, we will discuss the transition from Markovian to non-Markovian dynamics by tuning the coupling strength between the system and the non-Markovian bath.
%%%%%%%%%%%%%%%%%
%%%%%%%%%%%%%%%%%%%%%%
\section{Example: Non-Markovian dynamics of a harmonic oscillator in a dephasing environment}
\label{sec:III}
Our aim in this section is to substantiate the general discussion presented above using a particular example. To study the interplay between Markovian dynamics, we couple the system to both Markovian and non-Markovian baths, which also allows us to ensure the existence of a steady state. In fact,  from Eq.~\eqref{eq:stroboscopicVol}, we can see that if $|\det\hat{\Phi}(T)|<1$, the volume of accessible states stroboscopically and the system reaches a periodic steady state. In the example discussed in this section, the LO contains time-periodic rates. This can be achieved by engineering the non-Markovian bath, as we show in appendix~\ref{AppendixC}, where we propose an implementation of in circuit QED of the system discussed in this section.

%%%%%%%%%%%%%%%%%%%%%
\begin{figure}[t]
\centering
\includegraphics[scale=0.38]{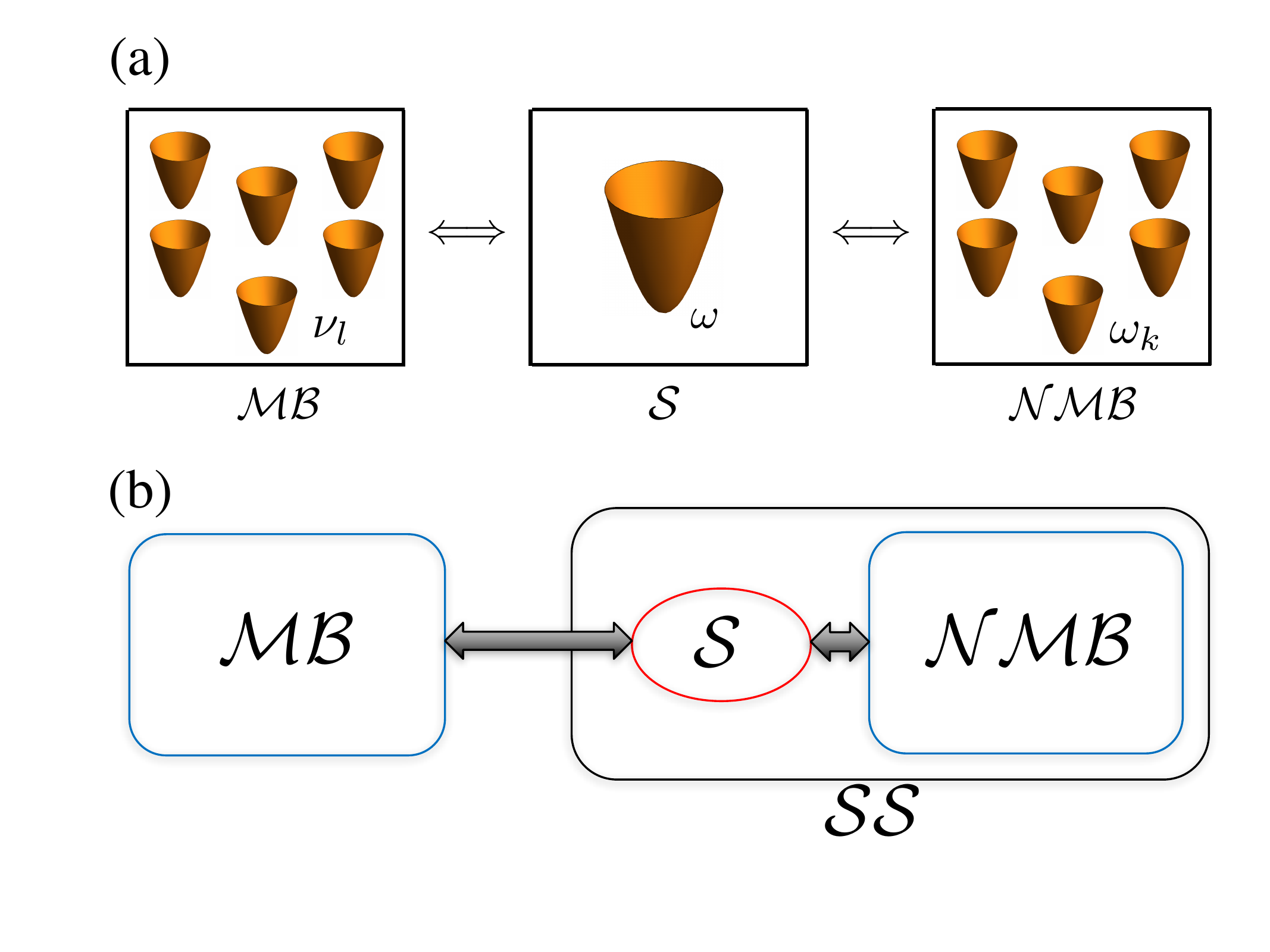}
\caption{a) Schematics representation of a harmonic oscillator $(\mathcal{S})$ of frequency $\omega$ interacting with a Markovian ($\mathcal{MB}$) and non-Markovian bath ($\mathcal{NMB}$). b) Sketch of the environment used for the microscopic derivation of the master equation.}
\label{fig:fig0}
\end{figure}
%%%%%%%%%%%%%%%%%%%%%%%
We begin by considering a time-local master equation for a harmonic oscillator coupled to two dephasing reservoirs, as shown in figure \ref{fig:fig0}(a). A natural way to derive the master equation is to consider the sketch of figure \ref{fig:fig0}(b) where we identify a super-system $\hat{H}_\mathcal{SS}=\hat{H}_{\mathcal{S}}+\hat{H}_{\mathcal{NMB}}+\hat{H}_{\mathcal{S}-\mathcal{NMB}}	$ composed of a harmonic oscillator $\hat{H}_\mathcal{S}=\omega \hat{n}$ coupled to a non-Markovian bath $\hat{H}_\mathcal{NMB}=\sum_{k=1}^{N_1} \omega_k \hat{b}^\dagger _k \hat{b}_k$ with $N_1$ modes, via the coupling Hamiltonian $\hat{H}_{\mathcal{S}-\mathcal{NMB}}=\hat{n}\sum_{k=1}^{N_1} g_k \left(\hat{b}^\dagger _k + \hat{b}_k \right)$.
In addition, the super-system is coupled to a Markovian bath $\hat{H}_\mathcal{MB}=\sum_{l=1}^{N_2} \nu_l \hat{c}^\dagger _l \hat{c}_l$ with $N_2$ modes via the interaction Hamiltonian $\hat{H}_{\mathcal{SS}-\mathcal{MB}}=\hat{n}\sum_{l=1}^{N_2} \eta_l \left(\hat{c}^\dagger _l + \hat{c}_l \right)$. The Hamiltonian of the total system is given by $\hat{H}=\hat{H}_\mathcal{SS}+\hat{H}_\mathcal{MB}+\hat{H}_{\mathcal{SS}-\mathcal{MB}}$.
We note that the operator $\hat{n}=\hat{a}^{\dagger}\hat{a}$ is defined in terms of bosonic operators $\hat{a}^{\dagger}$ and $\hat{a}$ of the harmonic oscillator. Correspondingly, $\hat{b}_k^{\dagger}$, $\hat{b}_k$ and $\hat{c}_l^{\dagger}$, $\hat{c}_l$ are bosonic operators of the non-Markovian and Markovian baths, respectively.

The master equation obtained from an exact microscopic derivation, see \ref{sec:MasterEq}, reads
%%%
\begin{equation}
      \label{eq:ResMastEq}
            \frac{d \hat{\rho}_{\mathcal{S}}(t)}{d t}=-i[\hat{H}_{\mathcal{S}}(t),\hat{\rho}_{\mathcal{S}}(t)]+ \gamma(t)\hat{\mathcal{D}}(\hat{n})\hat{\rho}_{\mathcal{S}}(t)
      \ ,
\end{equation}
%%%
where $
           \hat{\mathcal{D}}(\hat{O}_l)(\cdot)=\hat{O}_l(\cdot)\hat{O}_l^{\dagger}-\frac{1}{2}\{ \hat{O}_l^{\dagger}\hat{O}_l,(\cdot)\}
$.  We use units such that $\hbar=1$ and the coherent evolution is governed by the Hamiltonian $\hat{H}_{\mathcal{S}}(t)=\omega \hat{n}-g(t)\hat{n}^2$, where $\hat{n}=\hat{a}^{\dagger}\hat{a}$. Note that $\hat{H}_{\mathcal{S}} (t)$ is different from the original Hamiltonian of the system $\hat{H}_{\mathcal{S}}=\omega \hat{n} $, because it contains a Lamb shift $g(t)\hat{n}^2$ term that appears due to the interaction with the bath.
From Eq.~\eqref{eq:ResMastEq}, one can identify the structure of the Lindblad-type LO $\hat{\mathcal{L}}(t)(\cdot)=-i[\hat{H}_{\mathcal{S}}(t),(\cdot)] +\gamma(t)\hat{\mathcal{D}}(\hat{n})(\cdot)$~\cite{1976lindblad,1976gorini,2017Hartmann}. This master equation is motivated by previous works on phase damping~\cite{1985Walls} and dynamics of cavities coupled to moving mirrors~\cite{1997Bose}.
The total dephasing rate is given by (see \ref{sec:MasterEq3})
\begin{align}
\frac{\gamma(t)}{2}=\gamma_0+\sum^{N_1}_{k=1}\frac{g_{k}^2}{\omega_k}\sin(\omega_k t)\coth(\beta\omega_k/2),
\end{align}
where $\beta$ is the inverse temperature of the non-Markovian bath. The constant component $\gamma_0$ comes from the coupling to the Markovian bath. Besides its dephasing effect, the non-Markovian bath also influences the coherent dynamics of the system via the nonlinear driving strength $g(t)=\sum^{N_1}_{k=1}\frac{g_{k}^2}{\omega_k}(1-\cos\omega_k t)$. Without loss of generality, we consider a zero-temperature bath throughout the paper.

In order to have a time-periodic dephasing rate $\gamma(t)$ and driving $g(t)$, we consider a non-Markovian bath whose spectral density has peaks at frequencies $\omega_k=k^s\Omega$, where $s$ is a positive integer, and $\Omega=2\pi/T$.  For the purposes of this work, the bath frequencies are chosen as $\omega_k=k\Omega$ ($s=1$) with coupling strengths $g_k=he^{-zk/2}$, where $z>0$.  Interestingly, these requirements are almost natural in circuit QED setups. In Ref.~\cite{2017Gely}, for example, it is presented a microscopic description of a multimode resonator coupled to an artificial atom. In this implementation, the frequencies of the higher resonator modes are multiples of the fundamental frequency. Also, to avoid divergences of the Lamb-shift term, one has to introduce high-frequency cuttof by considering that higher modes will tend to decouple from the atom~\cite{2017Gely,2011Filipp}. We note that our results are valid for any value of $s$, and we take $s=1$ case for simplicity.  Also, our results are valid for any number $N_1$ of modes of the non-Markovian bath, even in the case of infinite numbers of modes, $N_1\to\infty$. Notice that our numerical calculations throughout the paper have been carried out  for a finite number of modes in the non-Markovian bath. In the case of a non-Markovian bath with infinite number of modes $N_1\rightarrow\infty$, the strength of bath-induced non-linearity is 
%%%
\begin{align}
            g(t)&=\sum^{\infty}_{k=1}\frac{g_k^2}{\omega_k}(1-\cos\omega_k t)
            \nonumber \\
            &=\frac{h^2}{\Omega}\left\{\text{Re}\left[\log(1-e^{-z+i \Omega t })\right]-\log(1-\cosh z+\sinh z)\right\}
\end{align}
%%%
and the dephasing rate reads
%%%
\begin{align}
\label{gammaT}
          \frac{\gamma(t)}{2} &=\gamma_0+\sum^{\infty}_{k=1}\frac{g_k^2}{\omega_k} \sin\omega_k t
            \nonumber \\
            &=\gamma_0-\frac{h^2}{\Omega}\text{Im}\left[\log(1-e^{-z+i \Omega t })\right]
            \ .
\end{align}
%%%

\subsection{Dynamics of the non-Markovian bath and stroboscopic divisibility}
\label{sec:MasterE4}
As we have an exact solution for the density matrix of the total system, we can explicitly calculate observables of the non-Markovian bath when $\gamma_0=0$ (in the absence of the Markovian bath). 
%%%%%%%%%%%%%%%%%%%%
\begin{figure}
\centering
\includegraphics[width=12.5cm,height=6.0cm]{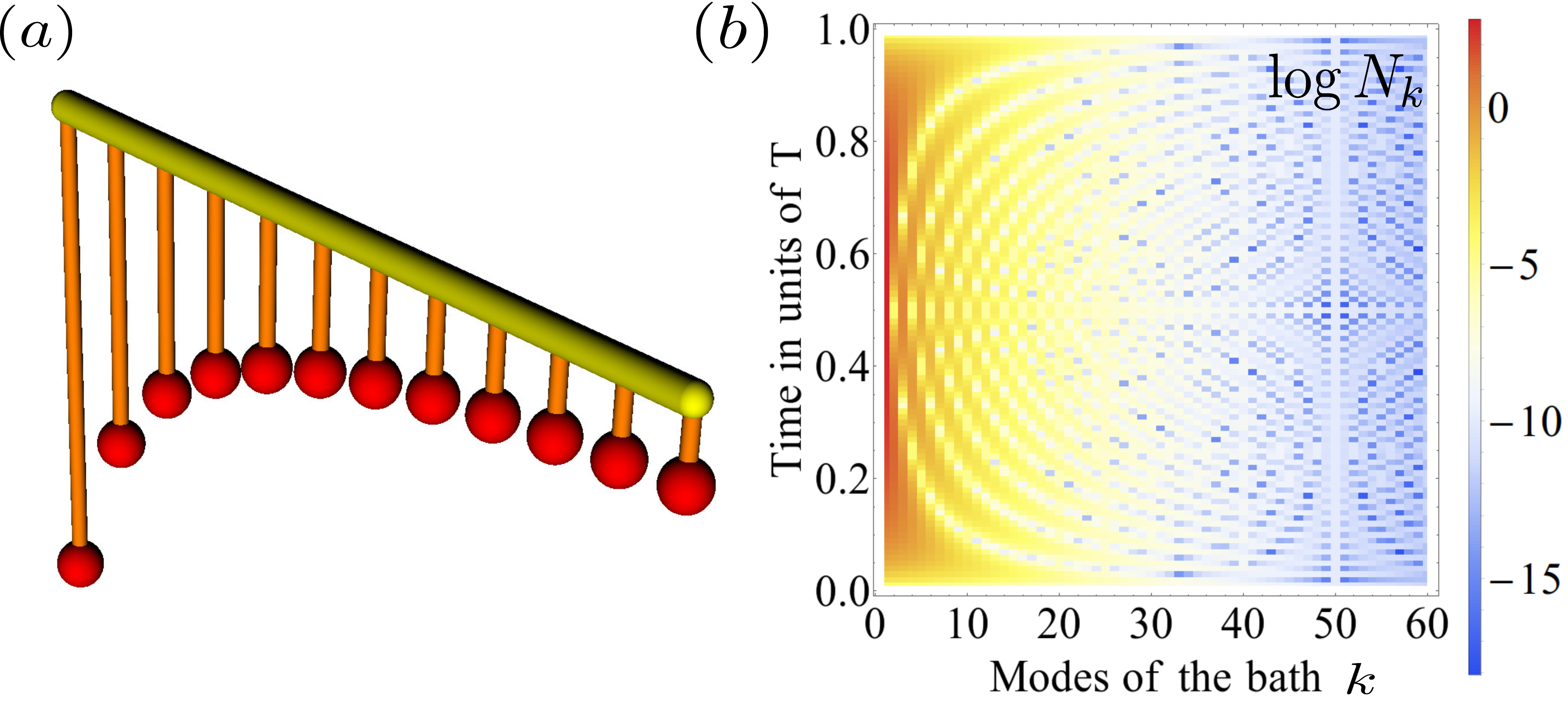}
\caption{Drawing of a pendulum wave device and dynamics of the mean photon number of a non-Markovian bath. (a) Depicts a device to demonstrate pendulum waves. In this mechanical device, the system comes back to its initial configuration after one time period $T$. (b) Quantum evolution of the mean photon number $N_k(t)=\expval{b^{\dagger}_kb_k(t)}$ of $N_1=60$ modes of the non-Markovian bath we consider in the manuscript (the density plot depicts $\log N_k$). Similarly to the pendulum waves, at a time $T$, the whole system comes back to its initial configuration. For the coupling $g_k=he^{-zk/2}$ to the modes of the bath we used $z=0.1$ and $h=1.0\Omega$. For convenience, we consider a zero temperature bath at the initial time with $N_k(0)=0$ and we prepare the resonator in a cat state $\ket{\Psi(0)}=C(\alpha_0)(\ket{\alpha_0}+\ket{-\alpha_0})$ with $|\alpha_0|=2$, where $C(\alpha_0)$ is a normalization factor. The frequency of the resonator is $\omega_0=10\Omega$.}
\label{fig:fig1}
\end{figure}

For example, for an initial state $\hat{\rho}_{\mathcal{S}}(0)=\ket{\Psi(0)}\bra{\Psi(0)}=\sum_{m,n}c_mc^{*}_n\ket{m}\bra{n}$, the mean photon number $N_k(t)=\expval{b^{\dagger}_kb_k(t)}$ for the $k$th mode reads
%%%
\begin{equation}
      \label{eq:kMeanPhotNum}
            N_k(t)=2\left(\frac{g_k}{\omega_k}\right)^2(1-\cos\omega_k t)\sum_{m}m^2|c_m|^2
      \ .
\end{equation}
%%%
Similarly, the expectation values of the quadratures $\hat{X}_k=\frac{1}{\sqrt{2}}(b^{\dagger}_k+b_k)$ and $\hat{P}_k=\frac{i}{\sqrt{2}}(b^{\dagger}_k-b_k)$ of the modes evolve as
%%%
\begin{align}
      \label{eq:kPosMean}
           \expval{\hat{X}_k(t)}&=\sqrt{2}\frac{g_k}{\omega_k}(\cos\omega_k t-1)\sum_{m}m|c_m|^2 \ ,
           \nonumber \\
           \expval{\hat{P}_k(t)}&=-\sqrt{2}\frac{g_k}{\omega_k}\sin\omega_k t\sum_{m}m|c_m|^2
      \ .
\end{align}
%%%
The physical intuition behind this solution is that the non-Markovian bath is out of equilibrium due to the coupling between the system and the bath and its time evolution is affected by the number of photons in the system. This is a total opposite to a Markovian dynamics, where the bath is not influenced by the system. Figure~\ref{fig:fig1}(a) depicts a mechanical analogue of the non-Markovian bath we are considering in the manuscript, which is referred to as pendulum waves~\cite{1991Pendulum}. One can prepare the ensemble of oscillators in a given configuration and after some time $T$ it will be back to the initial configuration. Similarly, figure~\ref{fig:fig1}(b) shows the dynamics of the bath with $N_1=60$ modes. To study dynamics, we initialize the system in a cat state $\ket{\Psi(0)}=C(\alpha_0)(\ket{\alpha_0}+\ket{-\alpha_0})$ with $|\alpha_0|=2$, where $C(\alpha_0)$ is a normalization factor. In this case, the period $T=2\pi/\Omega$ is determined by the fundamental frequency $\Omega$ and one can see that the dynamics of mean photon number of the modes $N_k(t)$ is reversed at time $t=T/2$, exactly as in the mechanical pendulum waves. This periodic motion of the non-Markovian bath is intimately related to the time-periodic rates, which allows us to define stroboscopic divisibility. 

As we can see from the previous discussion, our choice for the frequencies of the bath ($\omega_k=k^s\Omega$) has dramatic consequences for the time evolution of the system. In particular, from the expressions for the dephasing rate $\gamma(t)$ and the bath-induced nonlinearity $g(t)$, we find that these functions turn out to be periodic with period $T=2\pi/\Omega$. Besides this, the integral of the dephasing rate over one period is $\gamma_0>0$. An immediate consequence of this is that at times when the rates are positive, there is dephasing of the harmonic oscillator. Although the average of the rates in one period is positive, the rates can also be negative in certain intervals of time, where the coherences are built up again in the system. The latter is a signature of non-Markovianity~\cite{2017Vega, 2016Breuer, 2014Plenio}.
In \ref{AppendixC}, we propose an implementation of the system in circuit QED. 
%%%%%%%%%%%%%%%%%%%%%%
\section{Properties of the dynamical map}
\label{sec:IV}
The advantage of our exact solution for the master equation \eqref{eq:ResMastEq} is that the resulting dynamical map is valid for any strength of the coupling to the non-Markovian bath and for arbitrary spectral densities. For our choice of the bath frequencies, the Liouvillian is periodic and $|\det\hat{\Phi}(T)|\leq 1$. Based on the discussion of Eq.~\eqref{eq:stroboscopicVol}, one can see that the system is divisible at stroboscopic times  $t_l=lT$, where $l$ a positive integer. The physical interpretation of this is that the information trade-off between the system and the environment (Markovian plus Non-Markovian baths) is unbalanced and the volume of accessible states~\cite{2017Chruscinski,2014Maniscalco} is reduced stroboscopically. The information that goes away from the system when the rate is positive, is partially recovered if the rate becomes negative. A singular case of our results arises when the determinant of the dynamical map is one, i.e., the time average of the dephasing rate in one period is zero. In this situation, although one has non-Markovian dynamics, the system is purified stroboscopically and the discrete evolution is unitary.

%%%%%%%%%%%%%%%%%%%%%
\begin{figure}
\centering
\includegraphics[scale=0.5]{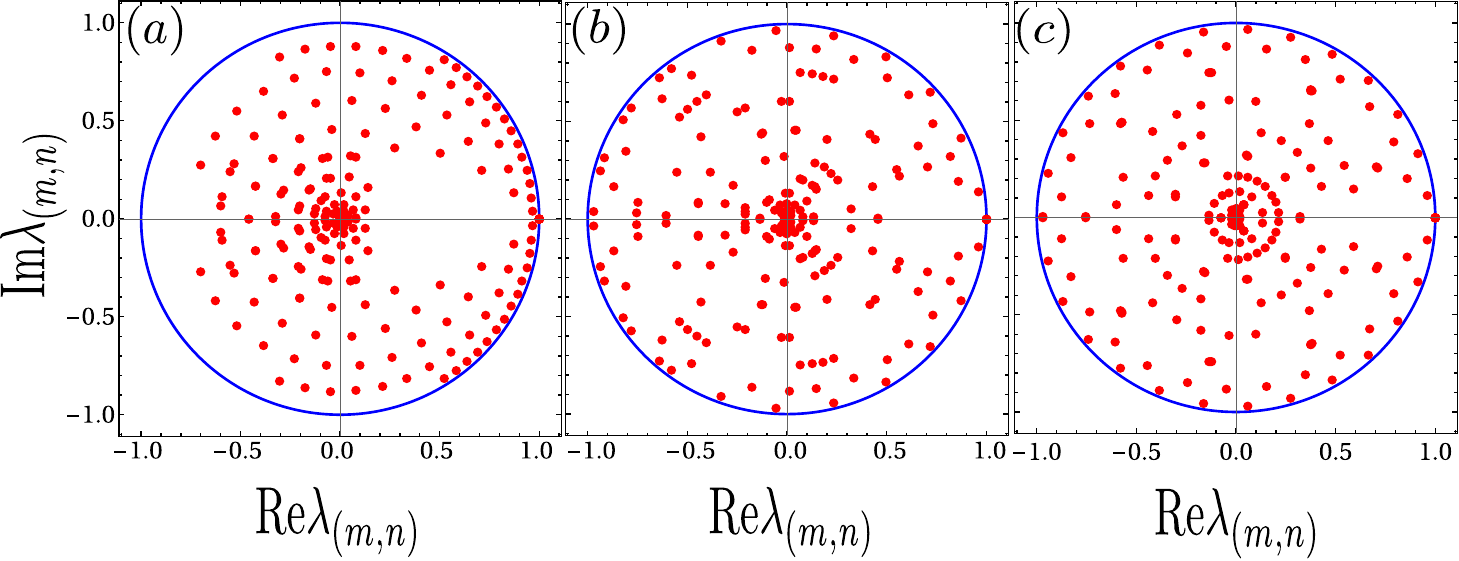}
\caption{Spectral properties of the dynamical map. We depict the characteristic multipliers $\lambda_{(m,n)}$ in the Markovian regime (a) $h=0.05\Omega$ and in the non-Markovian regime for (b) $h=0.1\Omega$ and (c) $h=1.0\Omega$. We truncated the Hilbert space of the resonator up to $n_t=14$ photons. We assume a coupling $g_k=he^{-zk/2}$ with $z=0.1$ between the system and the modes of the non-Markovian bath. We also consider a frequency $\omega=10\Omega$ of the resonator and the dephasing rate due to the Markovian bath is $\gamma_0=0.005\Omega$. We consider a non-Markovian bath with $N_1=60$ modes, but our results remain valid in the thermodynamic limit.}
\label{fig:fig2}
\end{figure}
%%%%%%%%%%%%%%%%%%%%

In our example, the dynamical map $\hat{\Phi}(T)$ is diagonal and its eigenvalues are the characteristic multipliers
%%%
\begin{equation}
      \label{eq:FloquetExp}
            \lambda_{m,n}=e^{-i(E_m-E_n)T}e^{iG(T)(m^2-n^2)}e^{-\gamma_0 T(m-n)^2}
\ ,
\end{equation}
%%%
where $E_n=n\omega$, $G(t)$ is the integral of the function $g(t)$, and $G(T)=T\sum_k\frac{g_k^2}{\omega_k}$ (see \ref{sec:MasterEq3}). From Eq.~\eqref{eq:FloquetExp} one can extract the Floquet-Liouville spectrum (Floquet exponents), because $L_{(n,m)}T=\log\lambda_{m,n}$  [we use the notation $\alpha=(m,n)$]. This information is of utmost importance because the real part of the Floquet exponents, i.e., the Lyapunov exponents, dictates the time to reach the steady state. In our case, this time scales as $1/\gamma_0$. The imaginary part of the Floquet-Liouville spectrum influences the coherent evolution of the system.

The characteristic multipliers can be depicted in the unit disk as shown in figure \ref{fig:fig2}. In the Markovian regime, where the rates are positive, we observe the clustering of the characteristic multipliers as depicted in figure \ref{fig:fig2}~(a). In the non-Markovian regime, the dephasing rate becomes negative in certain intervals~\cite{2017Vega, 2016Breuer, 2014Plenio, 2008Wolf,2010Rivas}. 
In this regime, we depict the characteristic multipliers in figures \ref{fig:fig2}(b) and (c). In contrast to figure \ref{fig:fig2}~(a), in the strong coupling regime $h^2/\Omega\gg \gamma_0$, the non-Markovian bath induces a nonlinearity proportional to $G(T)$, which is reflected in the repulsion of the eigenvalues as depicted in figure \ref{fig:fig2}(c).

%%%%%%%%%%%%%%%%%%%%%%%
\begin{figure}
\centering
\includegraphics[scale=0.5]{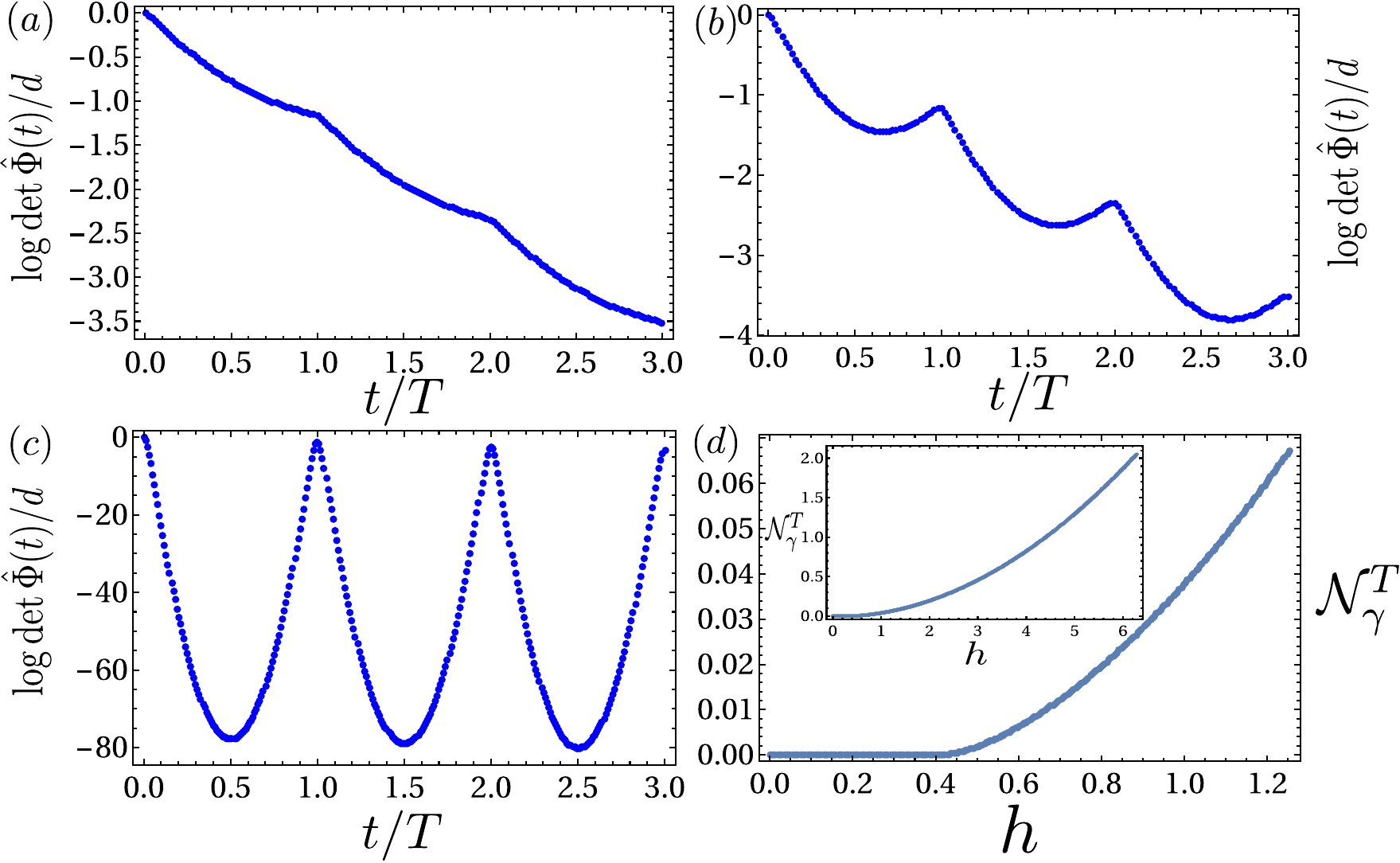}
\caption{Logarithm of the volume of accessible states $|\det\hat{\Phi}(t;0)|$ and non-Markovianity measure $\mathcal{N}^T_{\gamma}$. 
The panels $(a)$, $(b)$ and $(c)$ depict the $\log|\det\hat{\Phi}(t;0)|/d$ as a function of time for the same parameters as in figures \ref{fig:fig2}~$(a)$, $(b)$ and $(c)$, respectively. (d) Depicts the decay rate measure $\mathcal{N}^T_{\gamma}$ as a function of the coupling $h$ to the non-Markovian bath. One can see clearly the transition from Markovian $(\mathcal{N}^T_{\gamma}=0)$ to  non-Markovian $(\mathcal{N}^T_{\gamma}>0)$ dynamics. The inset depicts $\mathcal{N}^T_{\gamma}$ for higher values of $h$. Here $d=(n_t+1)^2$ , and $n_t=14$ is the truncation for the resonator Hilbert space.
All the parameters are the same as in figures \ref{fig:fig2}.} 
\label{fig:fig3}
\end{figure}
%%%%%%%%%%%%%%%%%%%%%%

\section{Non-Markovianity measure and dynamics of a Schr\"odinger cat state}
\label{sec:V}
We have  discussed so far spectral properties of the dynamical map. In this section, our aim is to present a quantification of non-Markovianity and its dynamical consequences. In the literature there are several measures of non-Markovianity~\cite{2014Plenio, 2016Breuer, 2017Vega, 2008Wolf,2010Rivas}. In our manuscript, we illustrate the general theory by considering an example. For convenience, we chose phase damping of an oscillator.  This leads to a master equation that has a single channel with decay rate $\gamma(t)$. As it is discussed in Ref.~\cite{2016Breuer}, in this case, the dynamical map is completely positive if the average of the decay rate is positive, i.e.,  $\int^{t}_{0}\gamma(\tau)d \tau>0$. In this case, however, although the average of the rates is positive, in intervals where the rates are negative, the map is not  CP-divisible~\cite{2016Breuer}.

On the other hand, our definition of \textit{Floquet stroboscopic divisibility} is based on the stroboscopic dynamics and it does not give us information about the non-Markovian behavior between two discrete times. One of the advantages of the example we are discussing in our manuscript is that the master equation~\eqref{eq:ResMastEq} has a single Lindblad operator $\hat{n}$ and in this case, all the different considered criteria for non-Markovianity coincide~\cite{2016Breuer}. Therefore, we decided to use one based on properties of the Liouvillian~\cite{2014Hall}, which is referred to as decay rate measure~\cite{2014Plenio}.  In our particular example, this measure is defined as $\mathcal{N}^T_{\gamma}=-\int^{t_b}_{t_a}\gamma(\tau)d \tau$.
The integration is carried out in the time interval $[t_a,t_b]$---within one period $T$, where the dephasing rate becomes negative. This measure is intimately related to the behavior shown in figures \ref{fig:fig3}~(a)-(c). There one can observe that in the Markovian case depicted in figure \ref{fig:fig3}~(a), the function $\log |\det\hat{\Phi}(t;0)|$ decreases monotonically and in the non-Markovian case of figures \ref{fig:fig3}~(b) and (c) it does not. In fact, the slope of the curves is proportional to $-\gamma(t)$. In the intervals where the slope becomes positive, the rates are negative, which is a signature of non-Markovian behavior. We depict the non-Markovianity measure in figure \ref{fig:fig3}~(d). There one can appreciate the transition between the Markovian and non-Markovian regimes as a function of the coupling to the non-Markovian bath.

%%%%%%%%%%%%%%%%%%%%%
\begin{figure}
\centering
\includegraphics[scale=0.25]{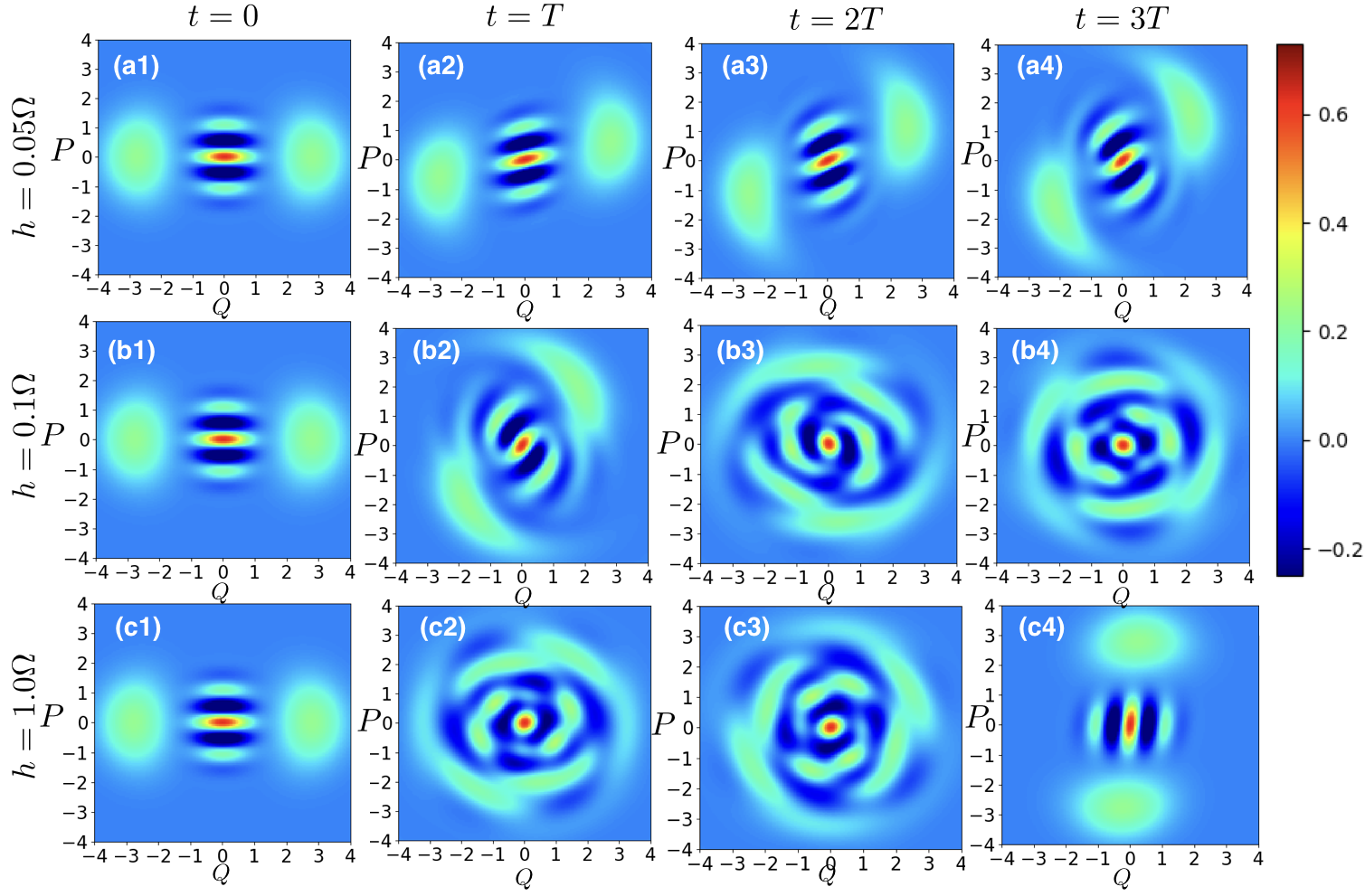}
\caption{Stroboscopic dynamics of the Wigner function at times $t_0=0$, $t_1=T$, $t_2=2T$, and $t_3=3T$. The panels (a1)-(a4) show  the evolution in the Markovian and (b1)-(b4), (c1)-(c4) in the non-Markovian regimes, respectively. All the parameters are the same as in figure \ref{fig:fig2}.} 
\label{fig:fig4}
\end{figure}
%%%%%%%%%%%%%%%%%%%%%%

Now let us explore the dynamical consequences of non-Markovian behavior. From the master equation (\ref{eq:ResMastEq}) one can see
that the non-Markovian bath introduces a time-dependent nonlinearity proportional to $g(t)\hat{n}^2$, which influences the coherent evolution of the harmonic oscillator. To study dynamics, let us suppose that the harmonic oscillator is initialized in a cat state $\ket{\Psi(0)}=C(\alpha_0)(\ket{\alpha_0}+\ket{-\alpha_0})$ with $|\alpha_0|=2$, where $C(\alpha_0)$ is a normalization factor. The initial density matrix is given by $\hat{\rho}_{\mathcal{S}}(0)=\ket{\Psi(0)}\bra{\Psi(0)}$.
To visualize the nonlinearity due to the coupling to the bath, we calculate the Wigner function $W(Q,P)=\frac{1}{\pi}\text{Tr}\left[\hat{\Pi}\ \hat{D}^{\dagger}(\alpha)\hat{\rho}_{\mathcal{S}}(t)\hat{D}(\alpha)\right]$ of the resonator, where $\hat{D}(\alpha)$  and $\hat{\Pi}$ are displacement and parity operators, respectively~\cite{1995Benedict}. By using the canonical coordinates $Q$ and $P$ one can define $\alpha=\frac{1}{\sqrt{2}}(Q+i P)$. The stroboscopic dynamics of the Wigner function is depicted in figures \ref{fig:fig4}~(a1)-(a4) in the Markovian case, and in figures \ref{fig:fig4}~(b1)-(b4) and (c1)-(c4) in the non-Markovian regime. When the system is strongly coupled to the non-Markovian bath, the Wigner function reveals signatures of the nonlinearity, as the system is not anymore in a cat state. However, after three periods of the evolution, the system is partially refocused to its initial state. The latter may be interpreted as a Poincar\'e recurrence since we are considering a finite number of modes in the non-Markovian bath.
%%%%%%%%%%%%%%%%%%%%%

%%%%%%%%%%%%%%%%%%%%%%
\section{Conclusions and outlook}
\label{sec:VI}
%%%%%%%%%%%%%%%%%%%%%
We have investigated the Liouvillian spectrum of a non-Markovian master equation which is local in time and has a periodic LO. Based on Floquet theory, we have shown that even though the dynamics is non-Markovian, the dynamical map is CP-divisible at stroboscopic times. In addition, we have proven that spectral properties of the LO determine the contraction of the volume of accessible states at stroboscopic times, which ensures the existence of a periodic steady state. To substantiate our theory, we present a time-local master equation derived microscopically for an environment composed of a non-Markovian and a Markovian bath.  We show that in this example, the volume of the accessible states ~\cite{2013Lorenzo} is stroboscopically reduced, because $|\det\hat{\Phi}(T)|<1$.  Possible directions in the future include the theoretical investigation of environments that exhibit phase transitions~\cite{2017cosco},  dissipative phase transitions \cite{PhysRevA.86.012116} and a time-nonlocal master equations~\cite{2017Magazzu}.

%%%%%%%%%%%%%%%%%%%%%%%%%%%%%%%%%%%%%
\section*{Acknowledgments}
V.M.B. and T.H.K. acknowledge fruitful discussions with A. Chia, I. Iakoupov, W. J. Munro,E. Munro, S. Restrepo, P. Strasberg, I. de Vega, and S. Vinjanampathy.The authors are grateful for the financial support through the National Research Foundation and Ministry of Education Singapore (partly through the Tier 3 Grant ``Random  numbers  from  quantum  processes''); and travel support by the EU IP-SIQS. The research leading to these results has received funding from the European Research Council under the European Union as Seventh Framework Programme (FP7/2007-2013) Grant Agreement No. 319286 Q-MAC and UK EPSRC funding EP/K038311/1. G.R. acknowledges the support from FONDECYT under grant No. 1150653. This research was also partially funded by Polisimulator
project co-financed by Greece and the EU
Regional Development Fund.%

\appendix

%%%%%%%%%%%%%%%%%%%%%%%%%%%%%%%%%%%%%%%%%%%%%%%%%%%%%%
\section{Microscopic derivation of the master equation}\label{microscopic1}
\label{sec:MasterEq}
In order to guide the reader through the microscopic derivation of the master equation (\ref{eq:ResMastEq}), we have divided this section in two subsections containing the steps of the derivation and its consequences. In subsection~\ref{Supersystem} we define the notation used in the derivation. In particular, we define the super-system which is composed by a harmonic oscillator coupled to a non-Markovian bath. The super-system is weakly coupled to a Markovian bath, which enables us to obtain a Lindblad-type master equation for the super-system reduced density matrix. In subsection~\ref{MasterEqResonator}, we trace out the degrees of freedom of the non-Markovian bath and give the explicit form the influence functional.  Once we have full knowledge of the reduced density matrix of the harmonic oscillator, one can obtain the master equation, as we describe at the end of the subsection.

\subsection{Derivation of the master equation for a super-system consisting of a harmonic oscillator plus non-Markovian bath\label{Supersystem}}
\label{sec:MasterEq1}
In this subsection we focus on the microscopic derivation of the master equation (\ref{eq:ResMastEq}) in the manuscript. Our derivation is based on the figure \ref{fig:fig0} of the manuscript. There we assume that the system is coupled to an environment which consist of two baths. One of them is Markovian and the other one is non-Markovian. A natural way to derive the master equation is to consider a super-system ($\hat{H}_\mathcal{SS}$) composed of the resonator ($\hat{H}_\mathcal{S}$) coupled to a non-Markovian bath ($\hat{H}_\mathcal{NMB}$) with $N_1$ modes via the coupling Hamiltonian $\hat{H}_{\mathcal{S}-\mathcal{NMB}}$. In addition, the super-system is coupled to a Markovian bath $\hat{H}_\mathcal{MB}$ with $N_2$ modes via the interaction Hamiltonian $\hat{H}_{\mathcal{SS}-\mathcal{MB}}$.
With the notation that we introduced in the figure \ref{fig:fig0}(b)  in the main text, we use the following Hamiltonians in the microscopic derivation
%%%
\begin{equation}
 \hat{H}=\hat{H}_\mathcal{SS}+\hat{H}_\mathcal{MB}+\hat{H}_{\mathcal{SS}-\mathcal{MB}}
 \ .	
\end{equation}
%%% 
As we discussed in the main text, there exists information flow back and forth between $\mathcal{S}$ and $\mathcal{NMB}$, due to the nature of the non-Markovian bath. This is possible because the system bath interaction
$H_{\mathcal{S}-\mathcal{NMB}}$ is not treated by perturbation theory. As a direct consequence of this, our treatment is valid for all the values of the coupling between the system and the non-Markovian bath. Notice that $H_{\mathcal{SS}-\mathcal{MB}}$ is the interaction between the super-system and the Markovian bath, that we consider to be in Born approximation with weak coupling, and we apply perturbation theory there.
In the following, we work in a frame where both $\mathcal{S}$ and $\mathcal{NMB}$ are diagonal. In so doing, we further transform all the Hamiltonian of the total system $\hat{H}$ into a polaron frame  and we represent it by using the superscript $p$. The polaron transformation is defined as
\begin{equation}
\label{eq:polaronTran}
	\hat{V}= \exp \left\lbrace \hat{n}\sum_k \frac{g_k}{\omega_k} \left(\hat{b}^\dagger _k -\hat{b}_k \right)\right\rbrace.
\end{equation}
In this new frame, we have $\hat{V}\hat{b}_k \hat{V}^{-1}=\hat{b}_k -\frac{g_k}{\omega_k}\hat{n}$, $\hat{V}\hat{n} \hat{V}^{-1}=\hat{n}$, and we define $\hat{H}^p=\hat{V}\hat{H} \hat{V}^{-1}$, where
\begin{eqnarray}
	\hat{H}^p &=& \hat{H}^p _{\mathcal{SS}} + \hat{H}^p _{\mathcal{MB}} + \hat{H}^p _{\mathcal{SS}-\mathcal{MB}}\nonumber\\
	&=& \underbrace{\omega_0 \hat{n}- \sum_{k=1}^{N_1} \frac{g_k ^2}{\omega_k}\hat{n}^2 + \sum_{k=1}^{N_1} \omega_k \hat{b}^\dagger _k \hat{b}_k }_{\hat{H}^p _{\mathcal{SS}}} + \underbrace{\sum_{l=1}^{N_2} \nu_l \hat{c}^\dagger _l \hat{c}_l }_{\hat{H}^p _{\mathcal{MB}}} + \underbrace{\hat{n}\sum_{l=1}^{N_2} \eta_l \left(\hat{c}^\dagger _l +\hat{c}_l \right)}_{\hat{H}^p _{\mathcal{SS}-\mathcal{MB}}}.
\end{eqnarray}

To derive a master equation for the reduced density matrix $\rho^p _\mathcal{SS} (t)$ of the super-system $\mathcal{SS}$, we assume that the super-system is weakly coupled to the Markovian bath.
We use then Born-Markov approximation~\cite{2002Breuer} to derive the master equation. As the direct consequence, the Markovian master equation for the system we consider (see figure \ref{fig:fig0} in the manuscript) is given by 
%%%
\begin{align}
      \label{eq:PolFrameMasterEq}
	\frac{d \hat{\rho}^p _\mathcal{SS} (t)}{dt}& =-i [\hat{H}^p _\mathcal{SS}, \hat{\rho}^p _\mathcal{SS} (t)]\nonumber\\
	&-\int_0 ^\infty d\tau e^{-i\hat{H}^p _{\mathcal{SS}}t} \text{Tr}_\mathcal{MB}\left\lbrace \left[\tilde{H}^p _{\mathcal{SS}-\mathcal{MB}}(t),\left[\tilde{H}^p _{\mathcal{SS}-\mathcal{MB}}(t-\tau),\tilde{\hat{\rho}}^p _\mathcal{SS}(t)\otimes \hat{\rho}^p _\mathcal{MB}\right] \right] \right\rbrace e^{i\hat{H}^p _{\mathcal{SS}}t} 
	\ ,
\end{align} 
%%%
where we have assumed that the density matrix $\tilde{\hat{\rho}}^p(t)$ of the total system, i.e., super-system plus Markovian bath, satisfies $\tilde{\hat{\rho}}^p(t)\approx \tilde{\hat{\rho}}^p _\mathcal{SS}(t)\otimes \hat{\rho}^p _\mathcal{MB}$. Here $\tilde{\hat{\rho}}^p _\mathcal{SS}(t)$ is a density matrix of the super-system and $\hat{\rho}^p _\mathcal{MB}=e^{-\frac{\hat{H}_\mathcal{MB}}{k_B T_{B}}}/\mathcal{Z}_\mathcal{MB}$ is a thermal density matrix of the Markovian bath. $\mathcal{Z}_\mathcal{MB}=\text{Tr}e^{-\frac{\hat{H}_\mathcal{MB}}{k_B T_{B}}}$ is the partition function and $T_{B}$ is the temperature of the Markovian bath.
Note that $\tilde{\hat{\rho}}^p(t)$ denotes the density matrix in the interaction picture and in the polaron frame.
Here, we use tilde sign to represent an operator is in its respective interaction picture. For instance, 
\begin{equation}
	\tilde{\mathcal{O}} (t)=\exp \left[i \left(\hat{H}^p _{\mathcal{SS}} + \hat{H}^p _{\mathcal{MB}}\right)t \right] \hat{\mathcal{O}} \exp \left[-i \left( \hat{H}^p _{\mathcal{SS}} + \hat{H}^p _{\mathcal{MB}}\right)t\right].
\end{equation}
When we expand equation \eqref{eq:PolFrameMasterEq}, we arrive at a simpler form:
\begin{align}
	\frac{d \hat{\rho}^p _\mathcal{SS} (t)}{dt}& =-i [\hat{H}^p _\mathcal{SS}, \hat{\rho}^p _\mathcal{SS} (t)]\nonumber\\
	&-\int_0 ^\infty d\tau \left\lbrace  \left[\hat{\rho}^p _\mathcal{SS}(t) e^{-i\hat{H}^p _\mathcal{SS} \tau}\hat{n} e^{i\hat{H}^p _\mathcal{SS} \tau}\hat{n}-\hat{n} \hat{\rho}^p _\mathcal{SS}(t) e^{-i\hat{H}^p _\mathcal{SS} \tau}\hat{n} e^{i\hat{H}^p _\mathcal{SS} \tau}\right]C(-\tau) \right. \nonumber \\
	&+\left. \left[\hat{n} e^{-i\hat{H}^p _\mathcal{SS} \tau}\hat{n} e^{i\hat{H}^p _\mathcal{SS} \tau} \hat{\rho}^p _\mathcal{SS}(t)-e^{-i\hat{H}^p _\mathcal{SS} \tau}\hat{n} e^{i\hat{H}^p _\mathcal{SS} \tau} \hat{\rho}^p _\mathcal{SS}(t) \hat{n} \right]C(\tau) \right\rbrace,\nonumber\\
\end{align}
where 
\begin{equation}
      \label{eq:bathcorrfunction1}
	C(t)= \expval{\tilde{\hat{B}}^p (t)\tilde{\hat{B}}^p(0)}= \text{Tr}_{\mathcal{MB}}[\tilde{\hat{B}}^p(t)\tilde{\hat{B}}^p(0)\hat{\rho}^p _\mathcal{MB}], \text{  and}
\end{equation}
\begin{equation}
	\tilde{\hat{B}}^p (t) = e^{i\hat{H}^p _{\mathcal{MB}}t} \left[ \sum_{l=1}^{N_2} \eta_l \left(\hat{c}^\dagger _l +\hat{c}_l \right)\right] e^{-i\hat{H}^p _{\mathcal{MB}}t}
	\ .
\end{equation}

Note that the expectation value of $\hat{x}_l=\sqrt{\frac{1}{2m\eta_l}}\left(\hat{c}^\dagger _l +\hat{c}_l \right)$ satisfies $\expval{\hat{x}_l}_\mathcal{MB}=\text{Tr}(\hat{x}_l\hat{\rho}^p _\mathcal{MB}) =0$. After some algebraic manipulations, we obtain

\begin{equation}
	{C(t)= \int_0 ^\infty d\omega \left[\cos(\omega t)\coth \left(\frac{\omega}{2k_B T_{B}}\right) -i\sin(\omega t) \right] J(\omega).}
\end{equation}
Following the same procedure as in Ref. \cite{1985Walls}, by taking the real part of $C(t)$, we can write down our Markovian master equation as
\begin{eqnarray}
      \label{eq:SuperSystemMasterEq}
	\frac{d \hat{\rho} _\mathcal{SS} (t)}{dt} &=&-i [\hat{H} _\mathcal{SS}, \hat{\rho} _\mathcal{SS} (t)]+ 2\gamma_0 \left[\hat{n}\hat{\rho} _\mathcal{SS}(t) \hat{n} -\frac{1}{2}\left\lbrace \hat{n}^2,\hat{\rho} _\mathcal{SS} (t)  \right\rbrace  \right],
\end{eqnarray}
where 
\begin{align}
	\gamma_0&=  \int_0 ^\infty d\tau \text{Re}[C(\tau)]=  \int_0 ^\infty d\tau \int_0 ^\infty d\omega \cos(\omega t)\coth \left(\frac{ \omega}{2k_B T_{B}}\right) J(\omega)\nonumber\\
	&=  \int_0 ^\infty d\omega \pi \delta(\omega)\coth \left(\frac{ \omega}{2k_B T_{B}}\right) J(\omega). 
\end{align}
We note that we dropped the superscript $p$ in the above master equation.  We did this because we transformed the master equation from the polaron frame into the original frame by performing the inverse polaron transformation defined in equation (\ref{eq:polaronTran}).

\subsection{Derivation of the final form of the master equation for the resonator\label{MasterEqResonator}}
\label{sec:MasterEq3}

In the interaction picture, the master equation \eqref{eq:SuperSystemMasterEq} has exact solution \cite{1985Walls} for the super-system density matrix
\begin{equation}
	\tilde{\hat{\rho}}_{\mathcal{SS}}(t)= \sum_{n,m} c_n c_m ^\ast e^{-\gamma_0 t(n-m)^2}\ketbra{n}{m}\otimes \hat{\rho}_\mathcal{NMB}(0).
\end{equation}
We have assumed that at the initial time, one has a factorized state of the super-system $\rho_\mathcal{SS}(0)=\rho_\mathcal{S}(0)\otimes \hat{\rho}_\mathcal{NMB}(0)$. At the initial time, the system is prepared in the state $\rho_\mathcal{S}(0)=\sum_{n,m} c_n c_m ^\ast \ketbra{n}{m}$.
We consider a non-Markovian bath which is in a thermal state $\hat{\rho} _\mathcal{NMB}(0)=e^{-\beta\hat{H}_\mathcal{NMB}}/\mathcal{Z}_\mathcal{NMB}$. Here $\mathcal{Z}_\mathcal{NMB}=\text{Tr}e^{-\beta\hat{H}_\mathcal{NMB}}$ is the partition function and $\beta$ is the inverse temperature.

In the Schr\"odinger picture, we have $\hat{\rho}_\mathcal{SS}(t)=e^{-i\hat{H}_\mathcal{SS}t}\tilde{\hat{\rho}}_\mathcal{SS}(t) e^{i\hat{H}_\mathcal{SS}t}$. In order to obtain the density matrix of the system (resonator), we need to trace out the degrees of freedom of the non-Markovian bath

\begin{align}
     \label{eq:exactResDenMat}
	\hat{\rho}_\mathcal{S}(t)&=\text{Tr}_{\mathcal{NMB}}[\hat{\rho}_\mathcal{SS}(t)]
	\nonumber\\&=
	\sum_{n,m} c_n c_m ^\ast e^{-\gamma_0 t(n-m)^2}\text{Tr}_{\mathcal{NMB}}\left[e^{-i\hat{H}_\mathcal{SS}t}\ketbra{n}{m}\otimes \hat{\rho}_\mathcal{NMB}(0)e^{i\hat{H}_\mathcal{SS}t}\right].
	\nonumber\\&=\sum_{n,m} c_n c_m ^\ast e^{-\gamma_0 t(n-m)^2}e^{-i(n-m)\omega_0 t}\ketbra{n}{m} F_{nm}(t)
	\ ,
\end{align}
%%%
where 
\begin{equation}
      \label{eq:NonMarkInflucenceFunc}
	F_{nm}(t)=\text{tr}\left(\hat{\rho}_\mathcal{NMB}(0) e^{i\hat{H}^{(m)}t} e^{-i\hat{H}^{(n)}t} \right),
\end{equation}
%%%
is the influence functional, 
and 
\begin{equation}
      \label{eq:FockStateHamiltonian}
	\hat{H}^{(n)}= \hat{H}_{\mathcal{NMB}} + n \sum_{k=1}^{N_1} g_k \left(\hat{b}_k ^\dagger +\hat{b}_k \right).
\end{equation}
Here, we note that $n$ is the eigenvalue of $\hat{n}$ in the Fock state basis of the resonator.
From these expressions one can see that the effect of the bath on the system is to create pure dephasing~\cite{1996Palma,2002Reina}. A related problem was discussed in the context on phase damping~\cite{1985Walls} and dynamics of cavities coupled to moving mirrors~\cite{1997Bose}. Now, the entire problem reduces to finding the influence functional $F_{nm}(t)$ analytically following a similar method as in Ref.~\cite{2002Reina}.  One can show that the influence functional is given by

\begin{equation}
	F_{nm}(t)= e^{i G(t)(n^2 -m^2)} e^{-\Gamma(t) (n-m)^2}
 \ ,
\end{equation}
where 
\begin{align}
      \label{eq:ParametersInfluenceFunc}
            G(t) &= \sum^{N_1}_{k=1} \left(\frac{g_k}{\omega_k} \right)^2 [\omega_k t -\sin (\omega_k t)], \text{\&}\\
	\Gamma (t) &= \sum^{N_1}_{k=1} \left(\frac{g_k}{\omega_k} \right)^2 [1-\cos (\omega_k t)] \coth \left(\frac{\beta \omega_k}{2}\right)
	\ .
\end{align}

By using the results obtained previously, we arrive at the exact solution for the resonator reduced density matrix
\begin{equation}
	\hat{\rho}_\mathcal{S}(t)=\Tr _\mathcal{NMB}[\hat{\rho}_\mathcal{SS}(t)]=\sum_{n,m} c_n c_m ^\ast e^{-i(n-m)\omega t} e^{-\gamma_0 t(n-m)^2} e^{i G(t) (n-m)^2} e^{-\Gamma(t) (n-m)^2} \ketbra{n}{m}.
\end{equation}
When we take the time derivative of the exact solution for $\hat{\rho}_\mathcal{S}(t)$, we obtain
\begin{eqnarray}
	\frac{d\hat{\rho}_{\mathcal{S}} (t)}{dt}
	&=& -i[\hat{H}_{\mathcal{S}} (t),\hat{\rho}_{\mathcal{S}} (t)]+ \gamma(t) \left[\hat{n}\hat{\rho}_{\mathcal{S}} (t) \hat{n}- \frac{1}{2}\left(\hat{n}^2 \hat{\rho}_{\mathcal{S}} (t) + \hat{\rho}_{\mathcal{S}} (t) \hat{n}^2 \right) \right]\nonumber\\
	&=& -i[\hat{H}_{\mathcal{S}} (t),\hat{\rho}_{\mathcal{S}} (t)]+ \gamma(t) \hat{\mathcal{D}}[\hat{n}]\hat{\rho}_{\mathcal{S}} (t).\label{first_master_eqn}
\end{eqnarray}
Here,
\begin{eqnarray}
	\hat{H}_{\text{S}} (t) &=& \omega \hat{n}- g(t) \hat{n}^2, \\
	g(t) &=& \dot{G}(t)=\sum^{N_1}_{k=1} \frac{g^2_k}{\omega_k}  [1 -\cos (\omega_k t)] ,\\
	\label{eq:nonLinDephasingRate}
	\gamma(t) &=& 2\gamma_0+ 2\dot{\Gamma}(t)=2\gamma_0+2\sum^{N_1}_{k=1} \frac{g^2_k}{\omega_k}  \sin (\omega_k t) \coth \left(\frac{\beta \omega_k}{2}\right)
	\ .
\end{eqnarray}
Based on this microscopic derivation, we have shown that the dephasing rate $\gamma(t)=2\gamma_0+2\dot{\Gamma}(t)$ can, in average, become non-zero when $\gamma_0> 0$.

%%%%%%%%%%%%%%%%%%%%%%%%%%%%%%%%%%%%%%
\section{Circuit QED implementation of the non-Markovian bath\label{AppendixC}}

\begin{figure}[ht]
\includegraphics[width=14cm,height=5cm]{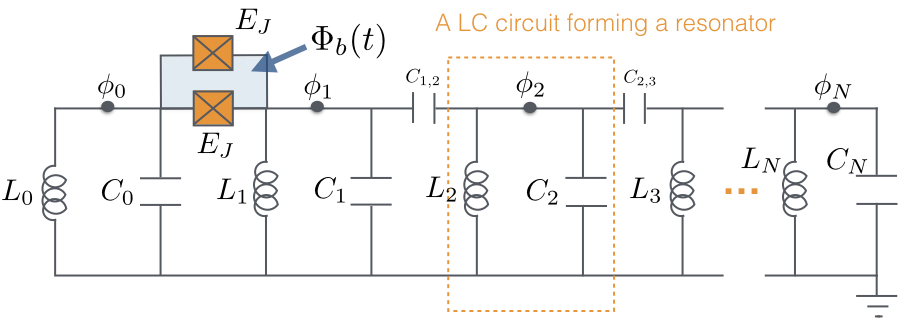}
\caption{Circuit diagram for implementing the system-bath Hamiltonian discussed in the main text.}
\label{fig:circuit}
\end{figure}

Circuit quantum electrodynamics (QED) has emerged as a promising platform to engineer strongly-correlated states of quantum matter, where ``particles" arise from excitations of low-temperature electrical circuits\cite{Houck:2012aa}. In this section we provide a circuit design that implements the Hamiltonian of the super system 
\begin{equation}
\hat{H}_{\mathcal{SS}} =\omega_0 \hat{a}^\dagger\hat{a} +\sum_{k=1}^N\omega_k \hat{b}^\dagger_k \hat{b}_k + \hat{a}^\dagger\hat{a}\sum_{k=1}^{N_1}g_k (\hat{b}^\dagger_k+\hat{b_k}).
\end{equation}
Here $\omega_0=\omega$ is the frequency of the resonator  in the main text.
To avoid nonlinear coupling between too many pairs of resonators, we apply the decomposition in eigenmodes $\hat{b}_k=\sum^{N_1}_{m=1} V_{k,m}\hat{\tilde{b}}_m$, where $V_{k,m}$ are the elements of a $N_1$ times $N_1$ square matrix. In terms of the new bosonic modes we obtain
\begin{equation}
\hat{H}_{\mathcal{SS}} =\omega_0 \hat{a}^\dagger\hat{a} + \sum_{m=1}^{N_1}\tilde{\omega}_m \hat{\tilde{b}}^\dagger_m \hat{\tilde{b}}_m + \sum_{m=1}^{N_1-1}J_m (\hat{\tilde{b}}^\dagger_m \hat{\tilde{b}}_{m+1} + \text{H.c.}) + \tilde{g}_1 \hat{a}^\dagger\hat{a}(\hat{\tilde{b}}^\dagger_1 + \hat{\tilde{b}}_1),
\label{eq:hsb}
\end{equation}
where the coupling $\tilde{g}_1=\sum^{N_1}_{k=1} V_{m,k}$ is related to the eigenmode decomposition discussed above. To substantiate the structure of the mode decomposition, we assume periodic boundary conditions $\hat{\tilde{b}}_1=\hat{\tilde{b}}^\dagger_{N_1+1}$ for the bosonic modes. This can be achieved by introducing a capacitive coupling between the nodes $1$ and $N_1$ in figure \ref{fig:circuit}. In this case, the coefficients appearing in the mode decomposition read $V_{m,k}=\frac{1}{\sqrt{N_1}}e^{i k m}$.

The latter equation is much less demanding as nonlinear coupling between only one pair of resonators is required. The circuit diagram for implementing the latter is shown in figure \ref{fig:circuit}. As will be shown below, each LC circuit forms a resonator with the frequency $\tilde{\omega}_m \approx 1/\sqrt{L_mC_m}$. These $LC$ circuits may as well be replaced with transmission lines which can be fabricated with higher precision \cite{PhysRevA.86.023837}. However, the calculation for the latter is more troublesome, so we restrict ourself to the $LC$ circuits instead for simplicity without compensating the physics. Nonlinear coupling comes from the use of the Josephson junctions with an external magnetic driving field.

Following the standard circuit quantization procedure \cite{DevoretLes}, we first write down the circuit's Lagrangian as 
\begin{align}
\mathcal{L}&=\sum_{m=0}^{N}\left(\frac{1}{2}C_m\dot{\phi}_m^2-\frac{1}{2L_m}\phi^2_m\right)+\sum_{m=1}^{N-1}\frac{1}{2}C_{m,m+1}(\dot{\phi}_m-\dot{\phi}_{m+1})^2 + E_J\cos\left(\frac{\phi_0-\phi_1+\Phi_b(t)}{\Phi_0}\right)\nonumber\\
&+ E_J\cos\left(\frac{\phi_0-\phi_1}{\Phi_0}\right),
\end{align}
where $C_m$, $C_{m,m+1}$ are capacitance, $L_m$ are inductance, $\Phi_0=\hbar/2e$ is the flux quantum, $E_J$ is the Josephson energy, $\Phi_b(t)=\pi+\phi_b(t)$ is a flux bias and $\phi_m=-\int V_m dt$ is a flux variable, with $V_m$ being a voltage at the corresponding position. We choose the flux bias field $\phi_b(t)$ to be an oscillating field with the frequency $\omega_p$, which can be implemented using an external AC magnetic field \cite{PhysRevLett.105.233907,PhysRevLett.119.150502}. The drive frequency will be chosen to eliminate undesired terms in the cosine expansion using the rotating wave approximation (RWA).

The Hamiltonian can be obtained by using the Legendre transformation,
\begin{align}
\hat{H}_{\mathcal{SS}}&=\sum_{m=0}^{N}\left(\frac{q_m^2}{2\tilde{C}_m}+\frac{\phi_m^2}{2L_m}\right)+\sum_{m=1}^{N-1}\frac{C_{m,m+1}}{\tilde{C}^2_m}q_mq_{m+1}-E_J\sum_{\eta=0}^{\infty}\frac{(-1)^\eta}{(2\eta)!}\left(\frac{\phi_0-\phi_1+\phi_b(t)}{\Phi_0}\right)^{2\eta}\nonumber\\
&-E_J\sum_{\eta=0}^{\infty}\frac{(-1)^\eta}{(2\eta)!}\left(\frac{\phi_0-\phi_1}{\Phi_0}\right)^{2\eta},
\label{eqApp:H}
\end{align}
where $q_m=\sqrt{\tilde{C}_m}\partial \mathcal{L}/\partial \dot{\phi_m}$ is a conjugate momentum of $\phi_m$, $\tilde{C}_m= C_{m,m-1}+C_{m,m+1}+C_m$ is an effective capacitance. Here we have assumed that $C_m/\tilde{C_m}\ll1$. We then quantized $\phi_m$ and $q_m$ by defining ladder operators $\hat{\tilde{b}}_m$, $\hat{\tilde{b}}^\dagger_m$ according to $\hat{\phi}_m  = (L_m/4\tilde{C}_m)^{1/4}(\hat{\tilde{b}}_m+\hat{\tilde{b}}^\dagger_m)$ and $\hat{q}_m = i(\tilde{C}_m/4L_m)^{1/4}(-\hat{\tilde{b}}_m+\hat{\tilde{b}}^\dagger_m)$. It follows that
\begin{align}
\sum_{m=0}^{N}\left(\frac{\hat{q}_m^2}{2\tilde{C}_m}+\frac{\hat{\phi}_m^2}{2L_m}\right) &= \sum_{m=0}^N\tilde{\omega}_m \hat{\tilde{b}}^\dagger_m\hat{\tilde{b}}_m,  \nonumber \\
\sum_{m=1}^{N-1}\frac{C_{m,m+1}}{\tilde{C}^2_m}\hat{q}_m\hat{q}_{m+1}&\approx \sum_{m=1}^{N-1}J_m(\hat{\tilde{b}}^\dagger_m\hat{\tilde{b}}_{m+1}+H.c.), \nonumber
\end{align}
where $\tilde{\omega}_m=1/\sqrt{L_m\tilde{C}_m}$ and $J_m=-\sqrt{\tilde{\omega}_m\tilde{\omega}_{m+1}} C_{m,m+1}/2\tilde{C}_m$. Here we have assumed that $J_m\ll \tilde{\omega}_m$ and hence the rotating term, $\hat{\tilde{b}}^\dagger_m\hat{\tilde{b}}^\dagger_{m+1}+H.c.$,  can be ignored with RWA. 

When working in the low excitation regime justified by a weak driving $\phi_b(t)\ll\tilde{\omega}_m$, the expansion of the cosine function term can be kept up to the fourth order ($\eta=0,1,2$). The quadratic term $(\eta=1)$ will simply renormalize the frequency of the resonators. This lefts us with only the fourth-order terms,
\begin{align}
&\frac{E_J}{12\Phi_0^4}\left[3\left(\hat{\phi}_0-\hat{\phi}_1\right)^{3}\phi_b(t)+3\left(\hat{\phi}_0-\hat{\phi}_1\right)^{3}\phi_b(t)+2\left(\hat{\phi}_0-\hat{\phi}_1\right)^{2}\phi_b(t)^2+\phi_b(t)^4\right] \nonumber \\
&\rightarrow -\frac{3E_J}{4\Phi_0^4} \hat{\phi}_0^2\hat{\phi}_1\phi_b(t) + .... \nonumber \\
&\rightarrow -\frac{ 3E_J}{4\Phi_0^4}\left(\frac{L_0^2L_1}{64\tilde{C}_0^2\tilde{C}_1}\right)^{1/4}\left((\hat{a}^\dagger)^2+\hat{a}^2+2\hat{a}^\dagger\hat{a}\right)(\hat{\tilde{b}}^\dagger_1+\hat{\tilde{b}}_1)\phi_b(t)+... .
\label{eq:nonlinear}
\end{align}
We then choose the coherent drive with the frequency $\omega_p=\tilde{\omega}_1$, i.e. $\phi_b(t) = \Omega(\hat{b}e^{-i \tilde{\omega}_1 t}+\text{H.c.})$ where $\hat{b}$ is promoted to a c-number. RWA can be applied for a weak driving $\Omega\ll\tilde{\omega}_m$. The only non-rotating term in equation \eqref{eq:nonlinear} that survives after the RWA is then $\hat{a}^\dagger\hat{a}(\hat{\tilde{b}}^\dagger_1+\hat{\tilde{b}}_1)$ as desired.
As we discussed before, by engineering the energies $\tilde{\omega}_m$ and couplings $J_m$, one can obtain a linear dispersion for the frequencies $\omega_k$ and the desired couplings $g_k$.

%% This BibTeX bibliography file was created using BibDesk.
%% http://bibdesk.sourceforge.net/

%% Saved with string encoding Unicode (UTF-8) 

\section*{References}

\providecommand{\newblock}{}

\end{document}